\documentclass[11pt,a4paper,nofootinbib,superscriptaddress,preprintnumbers]{revtex4-1}

\begin{document}

\preprint{YITP-17-63, IPMU13-0175}

\title{
Boulware-Deser ghost in extended quasidilaton massive gravity
}

\author{Shinji Mukohyama}

\affiliation{Center for Gravitational Physics, Yukawa Institute for Theoretical Physics, Kyoto University, 606-8502, Kyoto, Japan}
\affiliation{Kavli Institute for the Physics and Mathematics of the Universe (WPI), The University of Tokyo Institutes for Advanced Study, The University of Tokyo, Kashiwa, Chiba 277-8583, Japan}

\begin{abstract}
 In the extended quasidilaton massive gravity we perform a nonlinear
 transformation of the shift vector and then calculate the second
 derivatives of the Hamiltonian density with respect to the lapse
 function and the (nonlinearly transformed) shift vector. It is then
 shown that the $4\times 4$ Hessian matrix is invertible, meaning that
 the equations of motion for the lapse function and the shift vector
 simply determine themselves. Therefore, there is no primary constraint
 that removes the Boulware-Deser ghost. 
\end{abstract}

\maketitle

\section{Introduction}

The history of massive gravity dates back to 1939, when Fierz and Pauli
found the unique Lorentz-invariant linear theory without ghost. Since
then, there has been significant amount of progress on the subject,
including a number of important developments in early 1970's such as
discoveries of the vDVZ
discontinuity~\cite{vanDam:1970vg,Zakharov:1970cc}, the Vainshtein
mechanism~\cite{Vainshtein:1972sx} and the Boulware-Deser (BD)
ghost~\cite{Boulware:1973my}. However, it is only recent when a fully
nonlinear massive gravity without BD ghost, called dRGT theory, was
discovered~\cite{deRham:2010ik,deRham:2010kj}.

Having a good candidate theory of massive gravity, it is natural to ask
whether we can apply it to cosmology. Actually, it has been expected
that modification of gravity due to graviton mass may lead to cosmic
acceleration without dark energy. Unfortunately, it was 
reported that all homogeneous and isotropic cosmological solutions in
dRGT theory are unstable~\cite{DeFelice:2012mx}. Two options have thus
been suggested: (i) to change the background, or (ii) to change the
theory. The first option involves breaking of either
homogeneity~\cite{D'Amico:2011jj} or
isotropy~\cite{Gumrukcuoglu:2012aa,DeFelice:2013awa} of the background
configuration. The second option involves inclusion of extra degrees of 
freedom, either an extra scalar field~\cite{D'Amico:2012zv,Huang:2012pe}
or an additional spin-$2$ field~\cite{Hassan:2011zd,Hinterbichler:2012cn}.

Along the line of the second choice, the quasidilaton theory
originally proposed in \cite{D'Amico:2012zv} was extended by inclusion
of a new coupling between the massive graviton and the quasidilaton 
scalar~\cite{DeFelice:2013tsa}. In this extended quasidilaton theory,
five degrees of freedom of massive gravity as well as the quasidilaton
degree propagate on a strictly homogeneous and isotropic (FLRW),
self-accelerating de Sitter background and are stable. Both the new
coupling, which is denoted as $\alpha_{\sigma}$, and the canonical
kinetic term of the quasidilaton, whose coefficient is denoted as
$\omega$, are essential for the stability of degrees propagating on the
self-accelerating de Sitter solution: $\alpha_{\sigma}\ne 0$ provides
interaction that mixes the quasidilaton and the St\"uckelberg fields
while $\omega\ne 0$ provides interaction that mixes the quasidilaton and
the physical metric. In this fashion, all three parts can interact with
each other in a non-trivial way, and thus all together help evading the
no-go result of \cite{Gumrukcuoglu:2013nza,D'Amico:2013kya}.

The quasidilaton global symmetry allows for inclusion of Horndeski
terms~\cite{Horndeski:1974wa} with shift-symmetry as well. Those new
terms introduce further interaction between the quasidilaton and the
physical metric. Hence the shift-symmetric Horndeski terms may assist
or/and play the role of the $\omega$
term~\cite{DeFelice:2013dua}. Again, for any combinations of
shift-symmetric Horndeski terms (and the $\omega$ term), the inclusion
of the $\alpha_{\sigma}$ term is necessary for the stability of degrees
propagating on the self-accelerating FLRW de Sitter solution.

On the other hand, the DBI-type kinetic term (the term already suggested
in \cite{D'Amico:2012zv}, and the one proportional to the parameter
$\xi$ in the notation of \cite{Gumrukcuoglu:2013nza}) does not help
stabilizing the self-accelerating de Sitter
solution~\cite{DeFelice:2013dua} since it mixes the quasidilaton and the
St\"uckelberg fields via $\alpha_{\sigma}$ and thus does not induce the
kind of interaction that is provided by $\omega\ne 0$ (or the
shift-symmetric Horndeski terms). This has an important implication to
the relation between the extended quasidilaton massive gravity and the
DBI Galileon coupled to massive gravity (DBI massive
gravity)~\cite{Gabadadze:2012tr}: there is no overlap between these two
theories if we demand the stability of the self-accelerating FLRW de
Sitter solution~\cite{DeFelice:2013dua}.

One of the most important criteria for a consistent theory of massive
gravity is the absence of the BD ghost. At the level of linear
perturbations around the self-accelerating de Sitter solution in the
extended quasidilaton, it was explicitly shown in
\cite{DeFelice:2013tsa,DeFelice:2013dua} that there is no BD ghost, nor
any type of instabilities whose time-scales are parametrically shorter
than the cosmological time-scale. On the other hand, the argument for
the absence of BD ghost at fully nonlinear level so far relies on
arguments in the DBI massive gravity~\cite{Andrews:2013ora}. However, as
mentioned above, there is no overlap between the extended quasidilaton
and the DBI massive gravity if we demand the stability of the
self-accelerating FLRW de Sitter solution. For this reason, it is
necessary to study the issue of BD ghost at fully nonlinear level in 
the extended quasidilaton. It was shown by Kluson~\cite{Kluson:2013jea}
that a method that can find the primary constraint 
in the original quasidilaton theory does not apply to the extended
quasidilaton. This certainly shows difficulties in finding the primary 
constraint but actually does not prove either existence or non-existence
of the constraint. On the other hand, the argument in the 
previous arXiv version (v1) of the present paper was based on a wrong
gauge choice~\footnote{The condition $\phi^0=-e^{-\sigma/M_{\rm Pl}}$,
where $\phi^0$ is the temporal St\"uckelberg field and $\sigma$ is the
quasidilaton scalar, is not a gauge condition but a physical constraint.
Upon imposing this constraint, the system no longer describes the
extended quasidilaton theory. Instead, the system with the constraint
is one of Lorentz-violating and rotation-invariant massive gravity
theories.}, as pointed out in the second arXiv version (v2) of
\cite{Kluson:2013jea}. Thus it remains an open question whether the
extended quasidilaton theory possesses the primary constraint that
removes the BD ghost.

The purpose of the present paper (v2) is to settle the issue of the BD
ghost in the extended quasidilaton massive gravity. Actually, we shall
provide a proof of non-existence of the primary constraint that removes
the BD ghost. This means that the BD ghost exists in the extended
quasidilaton theory.

The rest of the paper is organized as follows. Sec.~\ref{sec:model}
briefly reviews the action of the extended quasidilaton massive
gravity. In Sec.~\ref{sec:newshift} we perform a nonlinear
transformation of the shift vector of the physical metric, using the
formalism developed in \cite{Hassan:2011tf}. In
Sec.~\ref{sec:minimalmodel}, in order to simplify the analysis we
set $\alpha_3=\alpha_4=1$.
In Sec.~\ref{sec:constraint} we then calculate the second
derivatives (i.e. components of the Hessian matrix) of the Hamiltonian
density with respect to the lapse and the (nonlinearly transformed)
shift vector. It is then shown that the Hessian matrix is
invertible. This implies that there is no primary constraint that
removes the BD ghost. Sec.~\ref{sec:summary} is devoted to a summary of
the paper.

\section{Extended quasidilaton}
\label{sec:model}

The action of the extended quasidilaton is given by 
\begin{equation}
S = \frac{M_{{\rm Pl}}^{2}}{2}\int d^{4}x\sqrt{-g}
 \left[
  R-2\Lambda
  -\frac{\omega}{M_{{\rm Pl}}^{2}}g^{\mu\nu}
  \partial_{\mu}\sigma\partial_{\nu}\sigma
  +2m_{g}^{2}(\mathcal{L}_{2}
  +\alpha_{3}\mathcal{L}_{3}+\alpha_{4}\mathcal{L}_{4})\right],
\end{equation}
where $\sigma$ is the quasidilaton scalar, $R$ is the Ricci scalar of
the physical metric $g_{\mu\nu}$ and the graviton mass terms are
specified as 
\begin{eqnarray}
\mathcal{L}_{2} & \equiv & \frac{1}{2}\,([\mathcal{K}]^{2}-[\mathcal{K}^{2}])\,,\nonumber\\
\mathcal{L}_{3} & \equiv & \frac{1}{6}\,([\mathcal{K}]^{3}-3[\mathcal{K}][\mathcal{K}^{2}]+2[\mathcal{K}^{3}])\,,\nonumber\\
\mathcal{L}_{4} & \equiv & \frac{1}{24}\,([\mathcal{K}]^{4}-6[\mathcal{K}]^{2}[\mathcal{K}^{2}]+3[\mathcal{K}^{2}]^{2}
+8[\mathcal{K}][\mathcal{K}^{3}]-6[\mathcal{K}^{4}])\,,
\end{eqnarray}
and
\begin{equation}
 \mathcal{K}_{\ \nu}^{\mu}=
  \delta_{\ \nu}^{\mu}
  -\left(\sqrt{g^{-1}\bar{f}}\right)_{\ \ \nu}^{\mu}\,.
\end{equation}
Here, $\bar{f}_{\mu\nu}$ is a combination of the Minkowski fiducial 
metric $\eta_{ab}=\mathrm{diag} (-1,1,1,1)$, the derivatives of the
St\"uckelberg fields $\partial_{\mu}\phi^a$ ($a=0,1,2,3$), the
quasidilaton $\sigma$ and its derivative $\partial_{\mu}\sigma$, 
specified as 
\begin{equation}
 \bar{f}_{\mu\nu} \equiv
  e^{2\sigma/M_{\rm Pl}}\eta_{ab}
  \partial_{\mu}\phi^a\partial_{\nu}\phi^b
  - \frac{\alpha_{\sigma}}{M_{\rm Pl}^2m_g^2}
  \partial_{\mu}\sigma\partial_{\nu}\sigma. 
  \label{eqn:barf}
\end{equation}
Note that $\bar{f}_{\mu\nu}$ defined here is related to
$\tilde{f}_{\mu\nu}$ introduced in \cite{DeFelice:2013tsa} as 
\begin{equation}
 \bar{f}_{\mu\nu} = e^{2\sigma/M_{\rm Pl}}\tilde{f}_{\mu\nu}. 
\end{equation} 
When $\alpha_{\sigma}=0$, the system reduces to the original
quasidilaton theory proposed in \cite{D'Amico:2012zv}.

The action, with or without the coupling $\alpha_{\sigma}$, enjoys the 
quasidilaton global symmetry, 
\begin{equation}
\sigma\to\sigma+\sigma_{0}\,,\qquad
 \phi^{a}\to e^{-\sigma_{0}/M_{{\rm Pl}}}\,\phi^{a}\,,
 \label{eqn:quasidilaton-symmetry}
\end{equation}
as well as the Poincare symmetry in the space of St\"uckelberg fields
\begin{equation}
\phi^{a}\to\phi^{a}+c^{a}\,,\qquad
 \phi^{a}\to\Lambda_{b}^{a}\phi^{b}\,.
\end{equation}

\section{Nonlinear transformation of shift vector}
\label{sec:newshift}

Let us adopt the ADM decomposition of the physical metric as
\begin{equation}
 g_{\mu\nu}dx^{\mu}dx^{\nu}
  = -N^2dt^2 + \gamma_{ij}(dx^i+N^idt)(dx^j+N^jdt), 
\end{equation} 
where $N$ ($>0$), $N^i$ and $\gamma_{ij}$ are the lapse function, the
shift vector and the spatial metric. It is also convenient to define $M$
($>0$), $M^i$ and $q_{ij}$ via the ADM decomposition of
$\bar{f}_{\mu\nu}$ as 
\begin{equation}
 \bar{f}_{\mu\nu}dx^{\mu}dx^{\nu}
  = -M^2dt^2 + q_{ij}(dx^i+M^idt)(dx^j+M^jdt). 
\end{equation} 
Concretely, 
\begin{equation}
 q_{ij} = \bar{f}_{ij}, \quad
  M^i = q^{ij}M_j, \quad
  M_i = \bar{f}_{0i}, \quad
 M^2 = -\bar{f}_{00} + M^kM_k,
\end{equation}
where $q^{ij}$ is the inverse matrix of $q_{ij}$.

Following \cite{Hassan:2011tf}, we perform a nonlinear field
redefinition from the original shift vector $N^i$ to a spatial
vector $n^i$ via the following relation. 
\begin{equation}
 N^i = n^i + M^i + ND^i_{\ j}n^j,
\end{equation} 
where $M^i$ is the shift vector of $\bar{f}_{\mu\nu}$ as defined above,
and $D^i_{\ j}$ is a matrix defined as follows. 
\begin{eqnarray}
& & D^i_{\ j} = \left(\sqrt{\gamma^{-1}qQ}\right)^i_{\ k}
 (Q^{-1})^k_{\ j},
\quad
Q^i_{\ j} = x\delta^i_{\ j} + n^in^kq_{kj}, \nonumber\\
& & (Q^{-1})^i_{\ j} = \frac{1}{x}(\delta^i_j
- M^{-2}n^in^kq_{kj}), \quad
x = M^2 - q_{kl}n^kn^l. \label{eqn:def-D}
\end{eqnarray} 
The matrix $D^i_{\ j}$ satisfies the following identities.
\begin{eqnarray}
& &  \sqrt{x}D =  \sqrt{(\gamma^{-1}-Dnn^TD)q}, \nonumber\\
& & q_{ik}D^k_{\ j} =  q_{jk}D^k_{\ i}, \quad
  D^i_{\ k}q^{kj} =   D^j_{\ k}q^{ki}, \quad
  q_{ik}(D^{-1})^k_{\ j} =   q_{jk}(D^{-1})^k_{\ i},
\end{eqnarray} 
where $(D^{-1})^i_{\ j}$ is the inverse matrix of $D^i_{\ j}$. It is
then shown that 
\begin{equation}
 N \left(\sqrt{g^{-1}\bar{f}}\right)^{\mu}_{\ \nu} = 
  A^{\mu}_{\ \nu} + NB^{\mu}_{\ \nu},
\end{equation}
where
\begin{eqnarray}
  \left(
   \begin{array}{cc}
    A^0_{\ 0} &  A^0_{\ j} \\
    A^i_{\ 0} &  A^i_{\ j}
   \end{array} 
       \right)
 & = & \frac{1}{\sqrt{x}} 
  \left(
   \begin{array}{cc}
    M^2 + n^kM_k & q_{jk}n^k\\
    -(M^2 + n^kM_k)(n^i+M^i)\  & \ -(n^i+M^i)q_{jk}n^k
   \end{array} 
       \right), \nonumber\\
  \left(
   \begin{array}{cc}
    B^0_{\ 0} &  B^0_{\ j} \\
    B^i_{\ 0} &  B^i_{\ j}
   \end{array} 
       \right)
  & = & \sqrt{x} 
  \left(
   \begin{array}{cc}
    0 & 0 \\
    D^i_kM^k\  & \ D^i_{\ j}
   \end{array} 
       \right).
\end{eqnarray}

What is important here is that $x$, $Q^i_{\ j}$, $(Q^{-1})^i_{\ j}$,
$D^i_{\ j}$, $A^{\mu}_{\ \nu}$ and $B^{\mu}_{\ \nu}$ are independent of
$N$ when components of $n^i$ (instead of $N^i$) are considered as
independent variables.

\section{Model with $\alpha_3=\alpha_4=1$}
\label{sec:minimalmodel}

Let us begin with fixing the gauge degrees of freedom as
\begin{equation}
 \phi^a =\delta^a_{\mu}x^{\mu}, \quad (a=0,1,2,3),
\end{equation} 
for which $\bar{f}_{\mu\nu}$ defined in (\ref{eqn:barf}) is 
\begin{equation}
 \bar{f}_{\mu\nu} = 
  e^{2\sigma/M_{\rm Pl}}\eta_{\mu\nu}
  -\tilde{\alpha}\partial_{\mu}\sigma\partial_{\nu}\sigma,
\end{equation}
where
\begin{equation}
 \tilde{\alpha} \equiv \frac{\alpha_{\sigma}}{M_{\rm Pl}^2m_g^2}. 
\end{equation} 
The ADM decomposition of $\bar{f}_{\mu\nu}$ then leads to 
\begin{eqnarray}
& & M = \tilde{M}\sqrt{1+\tilde{A}\dot{\sigma}^2}, \quad
  M_i = \tilde{M}_i\dot{\sigma}, \quad
  M^i = \tilde{M}^i\dot{\sigma}, \nonumber\\
& & 
q_{ij} = e^{2\sigma/M_{\rm Pl}}\delta_{ij}
  - \tilde{\alpha}\partial_i\sigma\partial_j\sigma, \quad
  q^{ij} = e^{-2\sigma/M_{\rm Pl}}
  \left(\delta^{ij}
  + \tilde{A}\delta^{ik}\delta^{jl}\partial_k\sigma\partial_l\sigma
  \right), 
\end{eqnarray}
where an overdot represents derivative with respect to $t$, and 
\begin{equation}
 \tilde{M} = e^{\sigma/M_{\rm Pl}}, \quad
  \tilde{M}_i = -\tilde{\alpha}\partial_i\sigma, \quad
  \tilde{M}^i = -\tilde{A}\delta^{ij}\partial_j\sigma, \quad
  \tilde{A} = \frac{\tilde{\alpha}e^{-\sigma/M_{\rm Pl}}}
  {1 - \tilde{\alpha}e^{-2\sigma/M_{\rm Pl}}
  \delta^{kl}\partial_k\sigma\partial_l\sigma}.
\end{equation}

Note that the dependence of the matrix $D^i_j$, defined in
(\ref{eqn:def-D}), on $\dot{\sigma}$ is nonlinear. To simplify the
dependence on $\dot{\sigma}$, we perform another field 
redefinition from $n^i$ to a new vector $\tilde{n}^i$ via
\begin{equation}
 n^i = M\tilde{n}^i,
\end{equation} 
and define $\tilde{x}$, $\tilde{Q}^i_j$ and $\tilde{D}^i_j$ by 
\begin{equation}
 x = M^2\tilde{x}, \quad Q^i_j=M^2\tilde{Q}^i_j, \quad
  D^i_j = \frac{1}{M}\tilde{D}^i_j,
\end{equation} 
so that
\begin{equation}
 N^i = M\tilde{n}^i + M^i + N\tilde{D}^i_{\ j}\tilde{n}^j
  = \tilde{M}\tilde{n}^i\sqrt{1+\tilde{A}\dot{\sigma}^2}
  + \tilde{M}^i\dot{\sigma}
  + N\tilde{D}^i_{\ j}\tilde{n}^j.
  \label{eqn:N-ntilde}
\end{equation} 
Hereafter, we consider $N$, $\tilde{n}^i$ (instead of $N^i$ or $n^i$),
$\gamma_{ij}$ and $\sigma$ as independent variables. It is then easy to
see that $\tilde{x}$, $\tilde{D}^i_{\ j}$, $\tilde{M}$ and $\tilde{M}^i$
are independent of $\dot{\sigma}$,  
\begin{equation}
 \frac{\partial \tilde{x}}{\partial \dot{\sigma}} = 0, \quad
  \frac{\partial \tilde{D}^i_{\ j}}{\partial \dot{\sigma}} = 0, \quad
  \frac{\partial \tilde{M}}{\partial \dot{\sigma}} = 0, \quad
  \frac{\partial \tilde{M}^i}{\partial \dot{\sigma}} = 0.
\end{equation} 
On the other hand, we have the following formulas for derivatives
with respect to $\tilde{n}^k$.
\begin{eqnarray}
& & \frac{\partial}{\partial \tilde{n}^k}\sqrt{\tilde{x}} = 
 -\frac{\tilde{n}_k}{\sqrt{\tilde{x}}}, \quad
 \frac{\partial}{\partial \tilde{n}^k}
 \left(\sqrt{\tilde{x}}\tilde{D}^i_{\ i}\right) = 
 -\frac{\tilde{n}_i}{\sqrt{\tilde{x}}}
  \frac{\partial}{\partial\tilde{n}^k}(\tilde{D}^i_{\ j}\tilde{n}^j), 
  \nonumber\\
& &  \frac{\partial}{\partial \tilde{n}^k}
 \left(\tilde{D}^i_{\ j}\tilde{n}^j\right) =
  \frac{1}{N}
  \left( \frac{\partial N^i}{\partial\tilde{n}^k}
   - \tilde{M}\delta^i_k\sqrt{1+\tilde{A}\dot{\sigma}^2}\right), 
  \quad
  \frac{\partial \tilde{M}}{\partial \tilde{n}^k} = 0, \quad
  \frac{\partial \tilde{M}^i}{\partial \tilde{n}^k} = 0,
\end{eqnarray}
where
\begin{equation}
 \tilde{n}_i = q_{ij}\tilde{n}^j.
\end{equation} 
One can also show from
$[\partial/\partial\tilde{n}^l,\partial/\partial\tilde{n}^k]
(\sqrt{\tilde{x}}\tilde{D}^i_{\ i})=0$ that
\begin{equation}
 \tilde{Q}_{ki}\frac{\partial N^i}{\partial\tilde{n}^l}
  = 
  \tilde{Q}_{li}\frac{\partial N^i}{\partial\tilde{n}^k},
  \label{eqn:QdNdn}
\end{equation} 
where
\begin{equation}
 \tilde{Q}_{ij} = q_{ik}\tilde{Q}^k_{\ j}
  = \tilde{x}q_{ij} + \tilde{n}_i\tilde{n}_j.
\end{equation}

For simplicity let us consider the case with $\alpha_3=\alpha_4=1$.
With this choice of parameters $\alpha_3$ and $\alpha_4$,
the graviton mass term is simplified as 
\begin{equation}
 \mathcal{L}_{2}+\mathcal{L}_{3}+\mathcal{L}_{4} 
  = -\left(\mathrm{tr}\sqrt{g^{-1}f}-3\right).
\end{equation}
By introducing momenta $\pi^{ij}$ conjugate to $\gamma_{ij}$ and
replacing $N^i$ with the r.h.s. of (\ref{eqn:N-ntilde}), the action
for the extended quasidilaton theory with $\alpha_3=\alpha_4=1$
is written as
\begin{equation}
 S_{\alpha_3=\alpha_4=1} = \int d^4x {\mathcal L}_{\alpha_3=\alpha_4=1},
\end{equation} 
where
\begin{eqnarray}
{\mathcal L}_{\alpha_3=\alpha_4=1} & = & \pi^{ij}\dot{\gamma}_{ij}
 + NR_0 + 
 \left(\tilde{M}\tilde{n}^i\sqrt{1+\tilde{A}\dot{\sigma}^2}
  +\tilde{M}^i\dot{\sigma}
  +N\tilde{D}^i_{\ j}\tilde{n}^j\right)R_i 
  - \frac{\omega}{2}\sqrt{\det\gamma}
  N\gamma^{ij}\partial_i\sigma\partial_j\sigma
  \nonumber\\
 & & + \frac{\omega\sqrt{\det\gamma}}{2N}
  \left[
   \dot{\sigma}
   -\left(\tilde{M}\tilde{n}^i\sqrt{1+\tilde{A}\dot{\sigma}^2}
     +\tilde{M}^i\dot{\sigma}
     +N\tilde{D}^i_{\ j}\tilde{n}^j\right)\partial_i\sigma
  \right]^2
 \nonumber\\
 & & 
  - M_{\rm Pl}^2m_g^2\sqrt{\det\gamma}
  \left(\tilde{M}\sqrt{\tilde{x}}\sqrt{1+\tilde{A}\dot{\sigma}^2}
   +N\sqrt{\tilde{x}}\tilde{D}^i_{\ i} - 3N\right),
\end{eqnarray} 
\begin{equation}
 R_0 = \frac{M_{\rm Pl}^2\sqrt{\det\gamma}}{2}\ {}^3R
  + \frac{1}{M_{\rm Pl}^2\sqrt{\det\gamma}}
  (\gamma_{ij}\gamma_{kl}-2\gamma_{ik}\gamma_{jl})\pi^{ij}\pi^{kl},
  \quad
  R_i = 2\gamma_{ik}\sqrt{\det\gamma}D_j\pi^{jk},
\end{equation}
${}^3R$ is the Ricci scalar of $\gamma_{ij}$ and $D_i$ is the covariant
derivative compatible with $\gamma_{ij}$.

\section{Non-existence of primary constraint}
\label{sec:constraint}

For the extended quasidilaton theory with $\alpha_3=\alpha_4=1$, the
canonical momentum conjugate to $\sigma$ is 
\begin{eqnarray}
 \pi_{\sigma} & = &
  \frac{\omega\sqrt{\det\gamma}}{N}
  \left[1-\left(\tilde{M}\tilde{n}^k
	   \frac{\tilde{A}\dot{\sigma}}{\sqrt{1+\tilde{A}\dot{\sigma}^2}}
	   +\tilde{M}^k\right)\partial_k\sigma\right] \nonumber\\
 & & \times
  \left[
   \dot{\sigma}
   -\left(\tilde{M}\tilde{n}^i\sqrt{1+\tilde{A}\dot{\sigma}^2}
     +\tilde{M}^i\dot{\sigma}
     +N\tilde{D}^i_{\ j}\tilde{n}^j\right)\partial_i\sigma
	\right] \nonumber\\
 & & 
  + \left(\tilde{M}\tilde{n}^i
	   \frac{\tilde{A}\dot{\sigma}}{\sqrt{1+\tilde{A}\dot{\sigma}^2}}
	   +\tilde{M}^i\right)R_i
  - M_{\rm Pl}^2m_g^2\sqrt{\det\gamma}\tilde{M}\sqrt{\tilde{x}}
  \frac{\tilde{A}\dot{\sigma}}{\sqrt{1+\tilde{A}\dot{\sigma}^2}}.
  \label{eqn:pi-sigma}
\end{eqnarray}
Hence we have
\begin{eqnarray}
 \frac{\partial\pi_{\sigma}}{\partial\tilde{n}^k}
  & = &
  -\frac{\omega\sqrt{\det\gamma}}{N}
  \left\{
   \frac{\tilde{M}\tilde{A}\dot{\sigma}\partial_k\sigma}
   {\sqrt{1+\tilde{A}\dot{\sigma}^2}}
   \left[
    \dot{\sigma}
    -\left(\tilde{M}\tilde{n}^i\sqrt{1+\tilde{A}\dot{\sigma}^2}
      +\tilde{M}^i\dot{\sigma}
      +N\tilde{D}^i_{\ j}\tilde{n}^j\right)\partial_i\sigma
   \right] \right. \nonumber\\
 & & 
  \left.
  +\left[1-\left(\tilde{M}\tilde{n}^k
	   \frac{\tilde{A}\dot{\sigma}}{\sqrt{1+\tilde{A}\dot{\sigma}^2}}
	   +\tilde{M}^k\right)\partial_k\sigma\right]
  \frac{\partial N^i}{\partial\tilde{n}^k}\partial_i\sigma
  \right\}
  + \frac{\tilde{M}\tilde{A}\dot{\sigma}}
  {\sqrt{1+\tilde{A}\dot{\sigma}^2}}R_k \nonumber\\
 & & 
  + M_{\rm Pl}^2m_g^2\sqrt{\det\gamma}\tilde{M}\tilde{n}_k
  \frac{\tilde{A}\dot{\sigma}}{\sqrt{1+\tilde{A}\dot{\sigma}^2}},
  \nonumber\\
 \frac{\partial\pi_{\sigma}}{\partial N} & = & 
  -\frac{\omega\sqrt{\det\gamma}}{N^2}
  \left[1-\left(\tilde{M}\tilde{n}^k
	   \frac{\tilde{A}\dot{\sigma}}{\sqrt{1+\tilde{A}\dot{\sigma}^2}}
	   +\tilde{M}^k\right)\partial_k\sigma\right] \nonumber\\
 & & \times
  \left[
   \dot{\sigma}
   -\left(\tilde{M}\tilde{n}^i\sqrt{1+\tilde{A}\dot{\sigma}^2}
     +\tilde{M}^i\dot{\sigma}\right)\partial_i\sigma
	\right], \nonumber\\
 \frac{\partial\pi_{\sigma}}{\partial \dot{\sigma}} & = & 
  \frac{\omega\sqrt{\det\gamma}}{N}
  \left[ (-1+\tilde{M}^k\partial_k\sigma)\tilde{M}\tilde{n}^l\partial_l\sigma
   \frac{\tilde{A}\dot{\sigma}(3+2\tilde{A}\dot{\sigma}^2)}
   {(1+\tilde{A}\dot{\sigma}^2)^{3/2}}
   + \tilde{A}(\tilde{M}\tilde{n}^k\partial_k\sigma)^2 
   \right. \nonumber\\
 & & \left.
   + (1-\tilde{M}^k\partial_k\sigma)^2 
   + N\tilde{D}^i_{\ j}\tilde{n}^j\partial_i\sigma
   \tilde{M}\tilde{n}^k\partial_k\sigma
   \frac{\tilde{A}}{(1+\tilde{A}\dot{\sigma}^2)^{3/2}}
  \right]  \nonumber\\
 & & 
  + \frac{\tilde{A}}{(1+\tilde{A}\dot{\sigma}^2)^{3/2}}
  (\tilde{M}\tilde{n}^iR_i
  -M_{\rm Pl}^2m_g^2\sqrt{\det\gamma}\tilde{M}\sqrt{\tilde{x}}),
  \label{eqn:derivative-of-pisigma}
\end{eqnarray} 
The Hamiltonian density is 
\begin{eqnarray}
 {\mathcal H}_{\alpha_3=\alpha_4=1} 
  & \equiv &
  \pi^{ij}\dot{\gamma}_{ij}+\pi_{\sigma}\dot{\sigma}
  - {\mathcal L}_{\alpha_3=\alpha_4=1}  \nonumber\\
  & = & -NR_0 
- \left(\frac{\tilde{M}\tilde{n}^i}{\sqrt{1+\tilde{A}\dot{\sigma}^2}}
   +N\tilde{D}^i_{\ j}\tilde{n}^j\right)R_i 
  + \frac{\omega}{2}\sqrt{\det\gamma}
  N\gamma^{ij}\partial_i\sigma\partial_j\sigma
  \nonumber\\
 & & + \frac{\omega\sqrt{\det\gamma}}{2N}
  \left\{
  \left[
   \dot{\sigma}
   -\left(\tilde{M}\tilde{n}^i\frac{\tilde{A}\dot{\sigma}^2}
     {\sqrt{1+\tilde{A}\dot{\sigma}^2}}
     +\tilde{M}^i\dot{\sigma}\right)\partial_i\sigma
  \right]^2
   \right.
 \nonumber\\
 & & 
  \left.\qquad\qquad\qquad
   -\left[\left(
	   \frac{\tilde{M}\tilde{n}^i}{\sqrt{1+\tilde{A}\dot{\sigma}^2}}
     + N\tilde{D}^i_{\ j}\tilde{n}^j\right)\partial_i\sigma\right]^2
  \right\} \nonumber\\
 & & 
  +M_{\rm Pl}^2m_g^2\sqrt{\det\gamma}
  \left(\frac{\tilde{M}\sqrt{\tilde{x}}}{\sqrt{1+\tilde{A}\dot{\sigma}^2}}
   +N\sqrt{\tilde{x}}\tilde{D}^i_{\ i} - 3N\right),
\end{eqnarray} 
where it is understood that $\dot{\sigma}$ is expressed implicitly in
terms of ($\pi_{\sigma}$, $N$, $\tilde{n}^k$, $\det\gamma$) via
(\ref{eqn:pi-sigma}).

Hereafter, we consider ($N$, $\tilde{n}^k$, $\gamma_{ij}$, $\pi^{\ij}$,
$\sigma$, $\pi_{\sigma}$) as independent variables. We are interested in
derivatives of the Hamiltonian density with respect to $N$ and
$\tilde{n}^k$. We define $\bar{\partial}/\bar{\partial}N$ and 
$\bar{\partial}/\bar{\partial}\tilde{n}^k$ to be partial derivatives
with respect to $N$ and $\tilde{n}^k$ with ($\tilde{n}^k$, 
$\gamma_{ij}$, $\pi^{\ij}$, $\sigma$, $\pi_{\sigma}$) and ($N$,
$\gamma_{ij}$, $\pi^{\ij}$, $\sigma$, $\pi_{\sigma}$), respectively,
fixed. In order to calculate these derivatives, we shall use 
(\ref{eqn:derivative-of-pisigma}) and the following formulas. 
\begin{equation}
 \frac{\bar{\partial}\dot{\sigma}}{\bar{\partial}N}
  = 
  -\left(\frac{\partial\pi_{\sigma}}{\partial N}\right)/
  \left(\frac{\partial\pi_{\sigma}}{\partial \dot{\sigma}}\right), \quad
 \frac{\bar{\partial}\dot{\sigma}}{\bar{\partial}\tilde{n}^k}
 =
 -\left(\frac{\partial\pi_{\sigma}}{\partial \tilde{n}^k}\right)/
 \left(\frac{\partial\pi_{\sigma}}{\partial \dot{\sigma}}\right), 
\end{equation}

The derivatives of ${\cal H}_{\alpha_3=\alpha_4=1}$ with respect $N$ and
$\tilde{n}^k$ are
\begin{eqnarray}
 {\cal H}_0  & \equiv & 
 \frac{\bar{\partial}{\cal H}_{\alpha_3=\alpha_4=1}}{\bar{\partial} N}
 = 
  \frac{\partial {\cal H}_{\alpha_3=\alpha_4=1}}{\partial N}
  - \frac{\partial {\cal H}_{\alpha_3=\alpha_4=1}}{\partial \dot{\sigma}}
  \left(\frac{\partial\pi_{\sigma}}{\partial N}\right)/
  \left(\frac{\partial\pi_{\sigma}}{\partial \dot{\sigma}}\right), 
  \nonumber\\
 & = & 
  -R_0 - \tilde{D}^i_{\ j}\tilde{n}^jR_i
  + M_{\rm Pl}^2m_g^2\sqrt{\det\gamma}
  \left(\sqrt{\tilde{x}}\tilde{D}^i_{\ i}-3\right) \nonumber\\
 & &
  + \frac{\omega\sqrt{\det\gamma}}{2N^2}
  \left\{ (\tilde{M}^i\partial_i\sigma-1)^2\dot{\sigma}^2
   +2(\tilde{M}^i\partial_i\sigma-1)\tilde{M}\tilde{n}^j\partial_j\sigma
   \dot{\sigma}\sqrt{1+\tilde{A}\dot{\sigma}^2} \right.
  \nonumber\\
 & & 
  \left.
   + \tilde{M}^2(1+\tilde{A}\dot{\sigma}^2)
   (\tilde{n}^i\partial_i\sigma)^2
   - N^2\gamma^{ij}\partial_i\sigma\partial_j\sigma
   - N^2\left(\tilde{D}^i_{\ j}\tilde{n}^j\partial_i\sigma\right)^2
  \right\}, \nonumber\\
 {\cal H}_k  & \equiv & 
  \frac{\bar{\partial}{\cal H}_{\alpha_3=\alpha_4=1}}{\bar{\partial}\tilde{n}^k}
  = 
  \frac{\partial {\cal H}_{\alpha_3=\alpha_4=1}}{\partial\tilde{n}^k}
  - \frac{\partial {\cal H}_{\alpha_3=\alpha_4=1}}{\partial \dot{\sigma}}
  \left(\frac{\partial\pi_{\sigma}}{\partial \tilde{n}^k}\right)/
  \left(\frac{\partial\pi_{\sigma}}{\partial \dot{\sigma}}\right) 
  = 
  {\cal C}_i\frac{\partial N^i}{\partial \tilde{n}^k}, \nonumber\\
 {\cal C}_i & = & -R_i - M_{\rm Pl}^2m_g^2\sqrt{\det\gamma}
  \frac{\tilde{n}_i}{\sqrt{\tilde{x}}} \nonumber\\
 & & 
  + \frac{\omega\sqrt{\det\gamma}}{N}
  \left[
   (1-\tilde{M}^k\partial_k\sigma)\dot{\sigma}
   -\tilde{M}\tilde{n}^k\partial_k\sigma\sqrt{1+\tilde{A}\dot{\sigma}^2}
   - N\tilde{D}^k_{\ l}\tilde{n}^l\partial_k\sigma
       \right] \partial_i\sigma, 
\end{eqnarray}
where it is again understood that $\dot{\sigma}$ is expressed implicitly
in terms of ($\pi_{\sigma}$, $N$, $\tilde{n}^k$, $\det\gamma$) via 
(\ref{eqn:pi-sigma}). Since the $3\times 3$ matrix 
$\partial N^i/\partial\tilde{n}^k$ is generically
invertible~\cite{Hassan:2011tf}, the set of equations of motion for $N$
and $\tilde{n}^k$ is equivalent to 
\begin{equation}
 {\cal H}_0 = 0, \quad {\cal C}_i = 0, \quad (i=1,2,3).
  \label{eqn:eom}
\end{equation}

In order to judge whether the equations of motion (\ref{eqn:eom})
completely determine $N$ and $\tilde{n}^k$ or leave some of them
undetermined, we need to know whether the Hessian matrix, whose
components are derivatives of ${\cal H}_0$ and ${\cal H}_k$ with respect
to $N$ and $\tilde{n}^l$, is invertible or not. Upon imposing
(\ref{eqn:eom}), these derivatives are calculated as
\begin{eqnarray}
 {\cal H}_{00}  & \equiv &
  \frac{\bar{\partial} {\cal H}_0}{\bar{\partial} N}
  = 
  \frac{\partial {\cal H}_0}{\partial N}
  - \frac{\partial {\cal H}_0}{\partial \dot{\sigma}}
  \left(\frac{\partial\pi_{\sigma}}{\partial N}\right)/
  \left(\frac{\partial\pi_{\sigma}}{\partial \dot{\sigma}}\right)
  = {\cal U}, \nonumber\\
 {\cal H}_{0l}  & \equiv & 
  \frac{\bar{\partial} {\cal H}_0}{\bar{\partial}\tilde{n}^l}
  =
  \frac{\partial {\cal H}_0}{\partial\tilde{n}^l}
  - \frac{\partial {\cal H}_0}{\partial \dot{\sigma}}
  \left(\frac{\partial\pi_{\sigma}}{\partial \tilde{n}^l}\right)/
  \left(\frac{\partial\pi_{\sigma}}{\partial \dot{\sigma}}\right)
  = -{\cal U}{\cal W} 
  \partial_i\sigma\frac{\partial N^i}{\partial\tilde{n}^l}, 
  \nonumber\\
 {\cal H}_{k0}  & \equiv &
  \frac{\bar{\partial} {\cal H}_k}{\bar{\partial} N}
  =
  \frac{\partial {\cal H}_k}{\partial N}
  - \frac{\partial {\cal H}_k}{\partial \dot{\sigma}}
  \left(\frac{\partial\pi_{\sigma}}{\partial N}\right)/
  \left(\frac{\partial\pi_{\sigma}}{\partial \dot{\sigma}}\right)
  = -{\cal U}{\cal W}
  \partial_i\sigma\frac{\partial N^i}{\partial\tilde{n}^k}, 
  \nonumber\\
 {\cal H}_{kl}  & \equiv & 
  \frac{\bar{\partial} {\cal H}_k}{\bar{\partial}\tilde{n}^l}
  =
  \frac{\partial {\cal H}_k}{\partial\tilde{n}^l}
  - \frac{\partial {\cal H}_k}{\partial \dot{\sigma}}
  \left(\frac{\partial\pi_{\sigma}}{\partial \tilde{n}^l}\right)/
  \left(\frac{\partial\pi_{\sigma}}{\partial \dot{\sigma}}\right)
  = {\cal V}\tilde{Q}_{ki}\frac{\partial N^i}{\partial\tilde{n}^l}
  + {\cal U}{\cal W}^2\partial_i\sigma\partial_j\sigma
  \frac{\partial N^i}{\partial\tilde{n}^k}
  \frac{\partial N^i}{\partial\tilde{n}^l}, 
  \label{eqn:H00H0kHk0Hkl}
\end{eqnarray}
where
\begin{equation}
 {\cal U} = \frac{\tilde{A}\gamma_0\gamma_1^2}{N^2\gamma_2}, 
  \quad
 {\cal V} = -\frac{M_{\rm Pl}^2m_g^2\sqrt{\det\gamma}}
  {\tilde{x}\sqrt{\tilde{x}}}, 
  \quad
 {\cal W} = \frac{N}{\gamma_1}, \label{eqn:def-UVW}
\end{equation}
and 
\begin{eqnarray}
 \gamma_0 & = & \omega M_{\rm Pl}^2m_g^2
  \tilde{M}\det\gamma, \quad
 \gamma_1 = \tilde{M}\tilde{n}^i\partial_i\sigma
  \sqrt{1+\tilde{A}\dot{\sigma}^2} + 
  (\tilde{M}^i\partial_i\sigma-1)\dot{\sigma}, 
  \nonumber\\
  \gamma_2 & = & \omega\sqrt{\det\gamma}\sqrt{\tilde{x}}
  \sqrt{1+\tilde{A}\dot{\sigma}^2}
  \left[\tilde{M}\tilde{n}^i\partial_i\sigma\tilde{A}\dot{\sigma}
  + (\tilde{M}^i\partial_i\sigma-1)\sqrt{1+\tilde{A}\dot{\sigma}^2}
  \right]^2 \nonumber\\
 & & 
  -NM_{\rm Pl}^2m_g^2\sqrt{\det\gamma}\tilde{M}\tilde{A}.
\end{eqnarray} 
Note that ${\cal H}_{0k}={\cal H}_{k0}$, 
${\cal H}_{kl}={\cal H}_{lk}$ (see (\ref{eqn:QdNdn})). 
Therefore, it is shown that
\begin{equation}
  \left(
   \begin{array}{cc}
    dN &  d\tilde{n}^k
   \end{array}
  \right)
  \left(
   \begin{array}{cc}
    {\cal H}_{00} & {\cal H}_{0l} \\
    {\cal H}_{k0} & {\cal H}_{kl}
   \end{array}
  \right)
  \left(
  \begin{array}{c}
   dN \\
   d\tilde{n}^l
  \end{array} 
       \right)
  =
  {\cal U}
  \left(dN+{\cal W}\partial_i\sigma
  \frac{\partial N^i}{\partial\tilde{n}^k}d\tilde{n}^k\right)^2
  + {\cal V}\tilde{Q}_{ki}\frac{\partial N^i}{\partial\tilde{n}^l}
  d\tilde{n}^kd\tilde{n}^l. \label{eqn:dNHdN}
\end{equation} 
As already mentioned before, the $3\times 3$ matrix 
$\partial N^i/\partial\tilde{n}^k$ is invertible. It is easy to
see that $\tilde{Q}_{kl}$ is also invertible as
\begin{equation}
 \left(\tilde{Q}^{-1}\right)^{kl} = 
  \frac{1}{\tilde{x}}\left(q^{kl}-\tilde{n}^k\tilde{n}^l\right). 
\end{equation}
Therefore, the form (\ref{eqn:dNHdN}) implies that the $4\times 4$
Hessian matrix is invertible whenever ${\cal U}{\cal V}\ne 0$. 
Note that ${\cal U}{\cal V}$ does not vanish generically if
$\omega\alpha_{\sigma}m_g^2\ne 0$. This means that the set of four 
equations (\ref{eqn:eom}) determines $N$ and $\tilde{n}^k$ ($k=1,2,3$).

After solving the set of four equations (\ref{eqn:eom}) w.r.t. $N$ and
$\tilde{n}^k$, the Hamiltonian density ${\mathcal H}_{\alpha_3=\alpha_4=1}$ is
expressed in terms of ($\gamma_{ij}$, $\sigma$, $\pi_{ij}$,
$\pi_{\sigma}$) only. Thus the Lagrangian density can be cast into the
first-order form (see e.g.~\cite{Faddeev:1988qp}) as 
\begin{equation}
 {\mathcal L}_{\alpha_3=\alpha_4=1} = 
  \pi^{ij}\dot{\gamma}_{ij}+\pi_{\sigma}\dot{\sigma}
  -  V(\gamma_{ij}, \sigma, \pi^{ij}, \pi_{\sigma}),
\end{equation} 
where $V(\gamma_{ij},\sigma,\pi_{ij},\pi_{\sigma})={\mathcal H}_{\alpha_3=\alpha_4=1}$. This explicitly shows that there is no primary constraint that
removes the Boulware-Deser ghost at a fully nonlinear level.

\section{Summary and discussions}
\label{sec:summary}

We have shown that the extended quasidilaton theory proposed in
\cite{DeFelice:2013tsa} does not have a primary constraint that removes 
the BD ghost at a fully nonlinear level, provided that
$\omega\alpha_{\sigma}m_g^2\ne 0$. Note that this is not just a failure
to find such a constraint but actually is a proof of non-existence. The
issue of BD ghost in the extended quasidilaton massive gravity has thus
been settled.

In the proof of non-existence of the primary constraint, we have assumed
that $\alpha_{\sigma}\ne 0$ and that $\omega\ne 0$. On the other hand, 
if $\alpha_{\sigma}=0$ or if $\omega=0$ then ${\cal U}$ defined in
(\ref{eqn:def-UVW}) vanishes and thus the Hessian matrix is not
invertible, meaning that there is a primary constraint. However, the
former choice $\alpha_{\sigma}=0$ does not allow for a stable
self-accelerating FLRW de Sitter
solution~\cite{DeFelice:2013tsa,DeFelice:2013dua}. In the later case
with $\omega=0$, one could introduce a DBI-type kinetic term for the
quasidilaton scalar (the term already suggested in
\cite{D'Amico:2012zv}, and the one proportional to the parameter $\xi$
in the notation of \cite{Gumrukcuoglu:2013nza}). The system will then
fall into a class of models proposed in
\cite{Gabadadze:2012tr}. (This case was considered also in
\cite{Kluson:2013jea}.) Unfortunately, this case again does not 
allow for a stable self-accelerating FLRW de Sitter
solution~\cite{DeFelice:2013dua}.

At the level of linear perturbations around the self-accelerating de
Sitter solution in the extended quasidilaton (with 
$\alpha_{\sigma}\ne 0$ and $\omega\ne 0$), it was explicitly shown in 
\cite{DeFelice:2013tsa,DeFelice:2013dua} that there is no BD
ghost. Hence, on the self-accelerating background, the BD ghost does not
show up at the linear level but should appear at some nonlinear level.
This is consistent with the recent claim of \cite{Anselmi:2017hwr} based
on linear perturbations around a more general FLRW background.

Having settled the issue of the BD ghost in the extended quasidilaton
massive gravity, we conclude that the extended quasidilaton
theory~\cite{DeFelice:2013tsa} is not a theoretically consistent candidate
for a theory of massive gravity. Therefore we need to consider other
candidates such as the minimal theory of massive gravity with two
propagating degrees of
freedom~\cite{DeFelice:2015hla,DeFelice:2015moy,DeFelice:2016ufg} and its
quasidilaton extension~\cite{DeFelice:2017wel}. The new quasidilaton
theory~\cite{Mukohyama:2014rca,DeFelice:2016tiu} is also another
possibility.

 \begin{acknowledgments}
The author thanks Antonio De Felice, Emir Gumrukcuoglu and Kazuya Koyama for useful comments. The work of the author was supported by Japan Society for the Promotion of Science (JSPS) Grants-in-Aid for Scientific Research (KAKENHI) No. 24540256, and by World Premier International Research Center Initiative (WPI), MEXT, Japan.
 \end{acknowledgments}

\end{document}